\documentclass[showpacs,amsmath,amssymb,aps]{revtex4} 
\usepackage{graphicx}
\usepackage{epstopdf}
\usepackage{amsmath}
\usepackage{bm}

\begin{document}

\title{Extra equation of gravity\\ induced by spontaneous local Lorentz symmetry breakdown}

\author{Kimihide Nishimura}
\email{kimihiden@dune.ocn.ne.jp}
\affiliation{Nihon-Uniform, 1-4-1 Juso-Motoimazato, Yodogawa-ku Osaka 532-0028 Japan}
\date{\today}

\begin{abstract}
A model of spontaneous Lorentz violation in four dimension is given, which seems to provide a Lorentz invariant effective theory. 
An SU(2) Yang-Mills gauge field and an auxiliary U(1) vector field generate gravity and other interactions 
when they have vacuum expectation values.  
The emergent gravity is equivalent to conventional general relativity up to the third order terms in the Lagrangian. 
The coupling to matter, including spin-1/2 fermions, is also given correctly to this level. 
It remains to be seen whether this formalism reproduces the properties of black holes and other consequences obtained from Einstein's theory.
\end{abstract}

\maketitle
\section{Introduction}
In contrast to broken gauge symmetry, broken spacetime symmetry has been rarely considered in four dimensions, 
since the latter is apt to end up with a Lorentz violating effective theory. 
However, this paper presents a model of spontaneous Lorentz violation in four dimensions, which seems to give a Lorentz invariant effective theory. 
We consider the four component gauge potential $Y_{\mu\alpha}$ composed of an SU(2) Yang-Mills gauge field  
and an auxiliary isoscalar vector field with the vacuum expectation value 
$\langle Y_{\mu\alpha}\rangle=\eta_{\mu\alpha}\phi_0$, where $\phi_0$ is a constant with mass dimension one. 
We can show that even if $Y_{\mu\alpha}$ couples to Weyl doublets, it is consistent with 
the relativity of emergent quasiparticles. 

Though the vacuum expectation value $\langle Y_{\mu\alpha}\rangle$ breaks both Lorentz symmetry and gauge symmetry, 
if isospin is rotated at the same time like a Lorentz vector, it remains invariant. 
In this case, the Minkowski metric $\eta_{\alpha\beta}$ should be assumed in the four-dimensional isospin space.
The invariance under the extended Lorentz transformation,  
$Y'_{\mu\alpha}=\Lambda_\mu{}^\nu \Lambda_\alpha{}^\beta Y_{\mu\beta}$ etc.,   
will therefore guarantee the Lorentz invariance of the effective theory. 

However, the above postulate is found not to be enough for the effective Lorentz invariance,  
since the mass term, the non-Abelian terms, and the interaction terms to fermions 
in the Lagrangian of the massive SU(2) Yang-Mills fields will not be invariant under the extended Lorentz transformation. 

The solution to this problem, on which this paper is mainly based, 
is to formulate the model in a spacetime with the quantum vierbein conformal to 
the gauge field, $\hat{e}_{\mu\alpha}=Y_{\mu\alpha}/\bar{\phi}$,  
where $\bar{\phi}$ is some scalar field with mass dimension one.  
The adjective ``quantum" is attached for distinguishing it from the classical vierbein in Einstein's theory of gravity. 
The new postulate returns to the original,  
when the quantum vierbein and the scalar field have the vacuum expectation values: 
$\langle\hat{e}_{\mu\alpha}\rangle=\eta_{\mu\alpha}$, and $\langle\bar{\phi}\rangle=\phi_0$, respectively.
Then, the gauge field $Y_\mu{}^\alpha$ provides the quantum mechanical local Lorentz frame vectors 
in the classical Minkowski spacetime,  
where the isospin index $\alpha$ is identified with that of the local Lorentz frame index. 
We then confirm the Lorentz invariance of the effective theory. 
Relativity of emergent fermions and their interactions are also confirmed. 

As a natural consequence of the above hypothesis, a graviton appears even in a flat Minkowski spacetime. 
Though the quantum metric $\hat{g}_{\mu\nu}$ obeys the Yang-Mills action for the massive SU(2) gauge field, 
the Einstein gravity is reproduced up to the third order terms in the Lagrangian. 
Whether the extra gravity alone is sufficient to completely replace Einstein gravity is an issue that this paper does not fully resolve.  

If there already exists Einstein's type gravity in unbroken phase, the extra gravity will become a renormalization.  
The phenomenological interpretations of the results obtained are finally discussed.
%
\section{Broken Lorentz symmetry and Goldstone bosons \label{BST&GB}}
As a preparation for understanding the necessity of quantum geometry by gauge fields for the effective Lorentz invariance, 
we begin with describing spontaneous Lorentz violation of a massive SU(2) gauge theory, 
and clarifying the number of resultant Nambu-Goldstone modes\cite{Goldstone,GSW}. 
The theme attracts attention also from the aspect of the breakdown of Goldstone's theorem\cite{LowManohar}. 

Spontaneous Lorentz violation occurs when the massive SU(2) gauge field $\bm{Y}_\mu$ couples to fermions.
We consider the Lagrangian ${\cal L}={\cal L}_Y+{\cal L}_F$ where
\begin{equation}
{\cal L}_Y=-\frac{1}{4}\bm{Y}^{\mu\nu}\cdot\bm{Y}_{\mu\nu}
+g\partial_\mu\bm{Y}_\nu\cdot(\bm{Y}^\mu\times\bm{Y}^\nu)
-\frac{g^2}{4}(\bm{Y}^\mu\times\bm{Y}^\nu)\cdot(\bm{Y}_\mu\times\bm{Y}_\nu)
+\frac{m_Y^2}{2}\bm{Y}^\mu\cdot\bm{Y}_\mu,
\label{Lag_Y}
\end{equation}  
$\bm{Y}_{\mu\nu}:=\partial_\mu\bm{Y}_\nu-\partial_\nu\bm{Y}_\mu$, and ${\cal L}_F$ is the Lagrangian of fermions, 
which will be specified in Sec.\ref{RQP}. 
The mass term is supposed to be originated from the Higgs mechanism, 
though the Lagrangian of the Higgs doublet is not explicitly written.  
In the Minkowski spacetime, the equation of motion for $\bm{Y}_\mu$ is given by 
\begin{equation}
(\square+m_Y^2)\bm{Y}_\mu-\partial_\mu\partial\cdot \bm{Y}+\cdots=\bm{j}_\mu, \quad 
 \bm{j}^\mu:=-\frac{\delta{\cal L}_F}{\delta\bm{Y}_\mu},
\label{EqM_Y}
\end{equation}
where the contributions from non-Abelian terms, which are not essential in the following argument, are neglected for simplicity. 
Then, the equation for the vacuum expectation value, 
\begin{equation}
m_Y^2\langle \bm{Y}_\mu\rangle=\langle \bm{j}_\mu\rangle,
\label{SCE}
\end{equation}
shows that $\langle \bm{Y}_\mu\rangle\neq0$, if $\langle \bm{j}_\mu\rangle\neq0$. 
When $\langle\bm{Y}_\mu\rangle\neq0$, the free equation of motion for a fermion depends on $\langle\bm{Y}_\mu\rangle$. 
In this case, the perturbative estimation gives $\langle\bm{j}_\mu\rangle=\Gamma(\bm{Y})\langle\bm{Y}_\mu\rangle$, 
where $\Gamma(\bm{Y})$ is generally a scalar function of $\langle\bm{Y}_\mu\rangle$ with mass dimension two. 
Then, Eq.(\ref{SCE}) provides the self-consistency condition for $\langle\bm{Y}_\mu\rangle$. 
If it has a nontrivial solution $\langle \bm{Y}_\mu\rangle\neq0$, spontaneous Lorentz violation will occur\cite{KN1,KN2}.  

We next consider this model in the Lagrangian formalism to find the Goldstone bosons. 
In the following, we assume that $\langle Y_{\mu a}\rangle=\eta_{\mu a}\phi_0$. 
Further, it is convenient for us to write the free part of the Lagrangian in the form  
\begin{equation}
\begin{array}{cc}
{\cal L}_Y^{(0)}=-\frac{1}{2}\partial^\rho Y^\mu{}_a\partial_\rho Y_{\mu a}+\displaystyle\frac{m_Y^2}{2}Y^\mu{}_aY_{\mu a},
\end{array}
\label{Lag'_Y}
\end{equation}
with the physical state condition 
\begin{equation}
\partial\cdot Y_a{}^{(+)}\rangle_{\rm phys}=0,
\label{PhysicalStateCondition}
\end{equation}
where $(+)$ denotes the positive frequency (annihilation) part as in the Gupta-Bleuler formalism \cite{Gupta,Bleuler}.
We can decompose $Y_{\mu a}$ into a local Lorentz transformation $\Lambda_\mu{}^\nu$,
and a symmetric tensor $\phi_{\mu a}$,   
\begin{equation}
Y_{\mu a}=\Lambda_\mu{}^\nu\phi_{\nu a},  
\end{equation}
where $\phi_{0a}=0$, $\phi_{ij}=\phi_{ji}$ and $\langle \phi_{\mu a}\rangle=\eta_{\mu a}\phi_0$. 
For a small local Lorentz transformation $\Lambda_\mu{}^\nu=\eta_\mu{}^\nu+\epsilon_\mu{}^\nu$,
the Lagrangian (\ref{Lag'_Y}) reduces to 
\begin{equation}
{\cal L}_Y=-\frac{\phi_0^2}{2}\partial^\rho\epsilon^\mu{}_ i\partial_\rho\epsilon_{\mu i}
+\frac{1}{2}\partial^\rho \phi_{ij}\partial_\rho \phi_{ij}
-\frac{m_Y^2}{2} \phi_{ij} \phi_{ij}. 
\label{Lag''_Y}
\end{equation}
Twelve components of $Y_{\mu a}$ are decomposed into the antisymmetric tensor $\epsilon_{\mu\nu}$ with six components, 
and the symmetric tensor $\phi_{ij}$ with also six components.
Six $\epsilon$ modes, corresponding to the Goldstone bosons due to Lorentz violation, are massless owing to the Lorentz invariance of the mass term $Y^\mu{}_aY_{\mu a}=\phi^\mu{}_a\phi_{\mu a}$.     

We next count the number of independent Goldstone bosons. 
By introducing the notations 
\begin{equation}
\begin{array}{cc}
\epsilon_{0i}=\epsilon_i,&\epsilon_{ij}=\epsilon_{ijk}\omega_k/\sqrt{2}, 
\end{array}
\label{notations}
\end{equation}
the Lagrangian is turned into the form
\begin{equation}
{\cal L}_Y=\frac{\phi_0^2}{2}\left[-\partial^\mu\epsilon_i\partial_\mu\epsilon_i+\partial^\mu\omega_i\partial_\mu\omega_i\right]
+\frac{1}{2}\partial^\rho \phi_{ij}\partial_\rho \phi_{ij}
-\frac{m_Y^2}{2} \phi_{ij} \phi_{ij}. 
\end{equation}
The canonical quantization in a volume $V$ gives the following expansion of field operators:
\begin{equation}
\begin{array}{cc}
\epsilon_i(x)\phi_0=\sum_{\bm{k}}\frac{1}{\sqrt{2\vert\bm{k}\vert V}}\left[a_{i\bm{k}}e^{-ikx}+a_{i\bm{k}}^\dagger e^{ikx}\right],&
\omega_i(x)\phi_0=\sum_{\bm{k}}\frac{1}{\sqrt{2\vert\bm{k}\vert V}}\left[b_{i\bm{k}}e^{-ikx}+b_{i\bm{k}}^\dagger e^{ikx}\right],
\end{array}
\label{ModeExp_epsilon}
\end{equation}
with the commutation relations:
\begin{equation}
\begin{array}{cc}
[a_{i\bm{k}}, a_{j\bm{k}'}^\dagger]=-\delta_{ij}\delta_{\bm{k}\bm{k}'},&
[b_{i\bm{k}}, b_{j\bm{k}'}^\dagger]=\delta_{ij}\delta_{\bm{k}\bm{k}'}.
\end{array}
\label{CR_epsilon}
\end{equation}
For the $\phi$ field, we introduce the notations 
$C_1=\sqrt{2}\phi_{23}$, $C_2=\sqrt{2}\phi_{31}$, $C_3=\sqrt{2}\phi_{12}$   
$D_1=\phi_{11}$, $D_2=\phi_{22}$, and $D_1=\phi_{33}$.  
The canonical quantization again gives the following field operators: 
\begin{equation}
\begin{array}{cc}
C_i(x)=\sum_{\bm{p}}\frac{1}{\sqrt{2\omega_{\bm{p}} V}}[c_{i\bm{p}}e^{-ipx}+c_{i\bm{p}}^\dagger e^{ipx}],&
D_i(x)=\sum_{\bm{p}}\frac{1}{\sqrt{2\omega_{\bm{p}} V}}[d_{i\bm{p}}e^{-ipx}+d_{i\bm{p}}^\dagger e^{ipx}],
\end{array}
\label{ModeExp_phi}
\end{equation}
with the commutation relations:
\begin{equation}
[c_{i\bm{p}}, c_{j\bm{p}'}^\dagger]=[d_{i\bm{p}}, d_{j\bm{p}'}^\dagger]=\delta_{ij}\delta_{\bm{p}\bm{p}'}.
\label{CR_phi}
\end{equation} 
Neglecting the infinite constant term, we obtain the Hamiltonian
\begin{equation}
H=\sum_{\bm{k},i}\vert\bm{k}\vert(-a_{i\bm{k}}^\dagger a_{i\bm{k}}+b_{i\bm{k}}^\dagger b_{i\bm{k}})
+\sum_{\bm{p},i}\omega_{\bm{p}}(c_{i\bm{p}}^\dagger c_{i\bm{p}}+d_{i\bm{p}}^\dagger d_{i\bm{p}}),
\end{equation}
where $\omega_{\bm{p}}=\sqrt{\bm{p}^2+m_Y^2}$. 
The vacuum state $\rangle:=\vert{\rm vacuum}\rangle$ is defined by the condition 
$(a_{i\bm{k}},b_{i\bm{k}},c_{i\bm{p}},d_{i\bm{p}})\rangle=0$. 
Although one-particle states $a_{i\bm{k}}^\dagger\rangle$ and $b_{i\bm{k}}^\dagger\rangle$ 
have the same positive energy eigenvalue $\vert\bm{k}\vert$, the former has the negative norm 
$\vert\vert a_{i\bm{k}}^\dagger\rangle\vert\vert^2=-1$, giving the negative energy expectation value 
$\langle a_{i\bm{k}}H a_{i\bm{k}}^\dagger\rangle=-\vert\bm{k}\vert$, while the state $b_{i\bm{k}}^\dagger\rangle$ with positive norm 
$\vert\vert b_{i\bm{k}}^\dagger\rangle\vert\vert^2=1$ gives the positive energy expectation value 
$\langle b_{i\bm{k}}H b_{i\bm{k}}^\dagger\rangle=\vert\bm{k}\vert$.

Because of the mass difference between the $\epsilon$ boson and the $\phi$ boson, 
the physical state condition (\ref{PhysicalStateCondition}) separates into two conditions:
\begin{equation}
\begin{array}{cc}
\partial^\mu\epsilon_{\mu a}^{(+)}\rangle_{\rm phys}=0,&
\partial^\mu \phi_{\mu a}^{(+)}\rangle_{\rm phys}=0.
\end{array}
\label{PhysicalStateCondition_epsilon_A}
\end{equation}
For an $\epsilon$ boson propagating in the direction of the third axis, $k^\mu=(\vert\bm{k}\vert,0,0,\vert\bm{k}\vert)$, 
the physical state condition reduces to 
\begin{equation}
\begin{array}{ccc}
(a_1+b_2/\sqrt{2})\rangle_{\rm phys}=0,&
(a_2-b_1/\sqrt{2})\rangle_{\rm phys}=0,&
a_3\rangle_{\rm phys}=0,
\end{array}
\label{PhysicalStateCondition_EpsilonPhi}
\end{equation}
where the momentum dependence is suppressed for simplicity. 
The vacuum state condition is expressible as $Z\rangle_{\rm phys}=0$,  
if the $Z$ operator is defined with constant coefficients $z_i$ by 
\begin{equation}
Z:=z_1(a_1+b_2/\sqrt{2})+z_2(a_2-b_1/\sqrt{2})+z_3a_3. 
\label{PhysicalStateConditionOperator}
\end{equation}
Then, we have the following relations:
\begin{equation}
\begin{array}{ll}
[Z, (a_1+\sqrt{2}b_2)^\dagger ]=0,&\vert\vert (a_1+\sqrt{2}b_2)^\dagger\rangle\vert\vert^2=1,\\  

[Z, (a_2-\sqrt{2}b_1)^\dagger ]=0,&\vert\vert (a_2-\sqrt{2}b_1)^\dagger\rangle\vert\vert^2=1,\\  

[Z, b_3^\dagger ]=0,&\vert\vert b_3^\dagger\rangle\vert\vert^2=1,
\end{array}
\label{PhysicalStateCondition_Z}
\end{equation}
which show that there are three independent physical states with positive norm and positive energy $\vert\bm{k}\vert$,  
since no other creation operators commuting with $Z$ are constructible from $a_i^\dagger$ and $b_i^\dagger$. 

For the $\phi$ boson, on the other hand, Eq. (\ref{PhysicalStateCondition_EpsilonPhi}) requires that
\begin{equation}
\begin{array}{ccc}
c_1\rangle_{\rm phys}=0,&
c_2\rangle_{\rm phys}=0,&
d_3\rangle_{\rm phys}=0,
\end{array}
\label{PhysicalStateCondition_phi}
\end{equation}
which shows that the $\phi$ boson has also three physical modes: 
$c^\dagger_3\rangle$, $d^\dagger_1\rangle$, and $d^\dagger_2\rangle$.

The number of Goldstone bosons ordinarily equals that of broken symmetries. 
However, we obtain here only three Goldstone bosons, although six Lorentz generators are broken. 
The phenomenon that the number of Nambu-Goldstone bosons becomes less than that of broken generators
for spontaneous spacetime symmetry breaking has been often reported, and the reason is known \cite{LowManohar}. 
We consider the variation of $Y_{\mu a}$ under an infinitesimal gauge transformation combined with a local Lorentz one:  
\begin{equation}
\delta Y_{\mu a}=\partial_\mu\theta_a+\epsilon_{abc}\theta_bY_{\mu c}+\epsilon_\mu{}^\nu Y_{\mu a},
\end{equation}
where $\theta_a$ are SU(2) gauge parameters. Then, we find that 
\begin{equation}
\begin{array}{cc}
\langle\delta Y_{0i}\rangle=\partial_0\theta_i+\epsilon_i\phi_0,&
\langle\delta Y_{ij}\rangle=\partial_i\theta_j+\epsilon_{ijk}(\omega_k/\sqrt{2}-\theta_k)\phi_0.
\end{array}
\end{equation}
The first relation shows that Lorentz boost transformations $\epsilon_i$ can be canceled by taking 
the gauge parameters such that $\partial_0\theta_i=-\epsilon_i\phi_0$, which implies that  
three Goldstone modes corresponding to the broken Lorentz boost transformations are not independent of gauge transformations. 
The second relation shows that, once gauge parameters $\theta_i$ are fixed to cancel the Lorentz boost modes, 
there remains no room for canceling the Goldstone modes corresponding to the broken rotational symmetry. 
Consequently, the total number of independent Goldstone modes for spontaneous violation of gauge and Lorentz symmetries are six: 
the three corresponding to SU(2) breaking, and the other three corresponding to rotational symmetry breaking.
Of those, the three gauge modes are to be absorbed as the longitudinal components of the massive gauge bosons 
by the Higgs mechanism. 
As a result, the final number of the Goldstone bosons is three, 
which is the same as that obtained from the physical state condition (\ref{PhysicalStateCondition}). 
\section{Extended relativity and Goldstone photon\label{GP}}
It is often anticipated that the photon may be the Goldstone boson emergent from spontaneous Lorentz violation 
\cite{Bjorken,Guralnik,Eguchi}. 
The last preparation is to show that two of the three Goldstone modes derived in the previous section are 
expressible in terms of a U(1) gauge field. 
The result serves to separate from the Lagrangian the part which 
breaks the extended Lorentz invariance. 

In order to acquire the perspectives on the relativistic properties of the effective theory, 
we introduce an auxiliary isoscalar gauge field $Y_{\mu0}$ with the vacuum expectation value 
$\langle Y_{\mu0}\rangle=\eta_{\mu0}\phi_0$ 
to form the four component gauge field $Y_{\mu\alpha}=(Y_{\mu0},Y_{\mu a})$   
with the vacuum expectation value, $\langle Y_{\mu\alpha}\rangle=\eta_{\mu\alpha}\phi_0$. 

We may assume $\langle Y_{\mu0}\rangle=\eta_{\mu0}\phi_0'$ instead of $\langle Y_{\mu0}\rangle=\eta_{\mu0}\phi_0$,  
where $\phi_0'$ is different from $\phi_0$. 
Even in this case, it can be changed into the original form by rescaling the time coordinate. 
Actually, by introducing the new coordinates $x'^0=cx^0$ and $x'^i=x^i$ with the ratio $c=\phi_0'/\phi_0$, we have
$\langle Y_\mu{}^0dx^\mu\rangle=cdx^0=\langle Y'_\mu{}^0dx'^\mu\rangle$, 
where $\langle Y'_\mu{}^0\rangle=\eta_\mu{}^0\phi_0$.
If $x^0$ has the dimension of time, $c$ represents the velocity of light.

The Lagrangian (\ref{Lag'_Y}) is expressible as 
\begin{equation}
{\cal L}_Y^{(0)}=\frac{1}{2}\partial^\rho Y^{\mu\alpha}\partial_\rho Y_{\mu\alpha}-\frac{1}{2}\partial^\rho A^\mu\partial_\rho A_\mu
+\frac{m_Y^2}{2}Y^\mu{}_aY_{\mu a}, 
\label{Lag_YA}
\end{equation}
with the physical state conditions $\partial\cdot Y_{\alpha}^{(+)}\rangle_{\rm phys}=0$, 
and $\partial\cdot A^{(+)}\rangle_{\rm phys}=0$, 
where $A_\mu:=Y_{\mu0}-\langle Y_{\mu0}\rangle$.  
The kinetic term for $Y_{\mu\alpha}$ in (\ref{Lag_YA}) becomes now invariant 
under the extended Lorentz transformation,  
\begin{equation}
Y'_{\mu\alpha}=\Lambda_\mu{}^\nu\Lambda_\alpha{}^\beta Y_{\nu\beta},
\end{equation}
where the four isospin components are also transformed in the same way as a Lorentz vector.
Under the same transformation, $A_\mu$ can be regarded as a Lorentz vector, $A'_\mu=\Lambda_\mu{}^\nu A_\nu$, 
if $A_\mu$ in an arbitrary Lorentz frame is defined by   
\begin{equation}
A_\mu=(Y_{\mu\alpha}-\langle Y_{\mu\alpha}\rangle)\delta^\alpha/\sqrt{\delta\cdot\delta},  
\end{equation}
with the help of a constant timelike 4-vector $\delta^\alpha$.
The 16 components of $Y_{\mu\alpha}$ are decomposable into a local Lorentz transformation $\Lambda_\mu{}^\nu$, 
and the residual symmetric tensor $\phi_{\mu\nu}=\phi_{\nu\mu}$ satisfying $\langle\phi_{\mu\alpha}\rangle=\eta_{\mu\alpha}\phi_0$ as 
\begin{equation}
Y_{\mu\alpha}=\Lambda_\mu{}^\nu\phi_{\nu\alpha}.
\end{equation}
For a small quantum oscillation, $\Lambda_\mu{}^\nu=\eta_\mu{}^\nu+\epsilon_\mu{}^\nu$, 
$Y_{\mu\alpha}$ is expressible up to the first order: 
\begin{equation}
Y_{\mu\alpha}=\phi_0\epsilon_{\mu\alpha}+\phi_{\mu\alpha}.
\end{equation}
Then, we have
\begin{equation}
\frac{1}{2}\partial^\rho Y^{\mu\alpha}\partial_\rho Y_{\mu\alpha}=
\frac{\phi_0^2}{2}\partial^\rho \epsilon^{\mu\alpha}\partial_\rho \epsilon_{\mu\alpha}
+\frac{1}{2}\partial^\rho \phi^{\mu\alpha}\partial_\rho \phi_{\mu\alpha}.
\end{equation}
The mass term is rewritten as
\begin{equation}
\frac{m_Y^2}{2}Y^\mu{}_aY_{\mu a}=\frac{m_Y^2}{2}\phi^\mu{}_a\phi_{\mu a}
=\frac{m_Y^2}{2}[\phi_{0i}^2-\phi_{ij}^2].
\label{MassTermForPhi}
\end{equation}
The fields $\epsilon_{\mu\nu}$, $\phi_{00}$, $\phi_{0i}$, and $\phi_{ij}$ obey the equations of motion:
\begin{equation}
\begin{array}{cccc} 
\square\epsilon_{\mu\nu}=0,&\square\phi_{00}=0, &[\square+\frac{m_Y^2}{2}]\phi_{0i}=0, &[\square+m_Y^2]\phi_{ij}=0. 
\end{array}
\end{equation}
Because of the mass difference, the physical state condition $\partial\cdot Y_\alpha^{(+)}\rangle_{\rm phys}=0$ 
separates into 
\begin{equation}
\begin{array}{cccc} 
\partial^\mu\epsilon_{\mu\nu}^{(+)}\rangle_{\rm phys}=0,&
\partial_0\phi_{00}^{(+)}\rangle_{\rm phys}=0,&
\partial_0\phi_{0i}^{(+)}\rangle_{\rm phys}=0,&
\partial_j\phi_{ji}^{(+)}\rangle_{\rm phys}=0.
\end{array}
\label{PhysicalStateCondition2}
\end{equation}
As for the first condition, 
the expressions (\ref{ModeExp_epsilon}) and (\ref{CR_epsilon}) become applicable also to the present case 
by taking $\epsilon_{0i}=\epsilon_i/\sqrt{2}$ instead of $\epsilon_{0i}=\epsilon_i$ in Eq.(\ref{notations}).
The $Z$ operator corresponding to Eq.(\ref{PhysicalStateConditionOperator}) is now written as 
\begin{equation}
Z:=z_1(a_1+b_2)+z_2(a_2-b_1)+z_3a_3,
\label{PhysicalStateConditionOperator2}
\end{equation}
Then, we have
\begin{equation}
\begin{array}{ll}
[Z, (a_1+b_2)^\dagger ]=0,&\vert\vert (a_1+b_2)^\dagger\rangle\vert\vert^2=0,\\  

[Z, (a_2-b_1)^\dagger ]=0,&\vert\vert (a_2-b_1)^\dagger\rangle\vert\vert^2=0,\\  

[Z, b_3^\dagger ]=0,&\vert\vert b_3^\dagger\rangle\vert\vert^2=1, 
\end{array}
\label{PhysicalStateCondition_R}
\end{equation}
which shows that only $b_3^\dagger\rangle$ remains as the Goldstone boson, 
and the other two states in the previous section turn into zero-norm states carrying no energy.
Instead of two Goldstone modes lost, we now obtain a U(1) gauge field $A_\mu$ with two polarization modes.
This observation is interpretable as that two of the three Goldstone bosons in the first formulation are transformed 
into a massless U(1) gauge boson in the second formulation. 
The result justifies one to treat $Y_{\mu0}$ and $A_\mu$ in (\ref{Lag_YA}) as independent.

Concerning the $\phi$ boson, on the other hand, 
owing to the second and the third condition in (\ref{PhysicalStateCondition2}), 
$\phi_{00}$ component and $\phi_{0i}$ components are physically prohibited.
Only the third condition remains, which is the same as that for $\phi_{ij}$ in the previous section. 
Therefore, the number of physical modes for the $\phi$ boson is the same also in the new formulation. 

The new formulation improves the perspectives on the Lorentz symmetry of the effective theory, 
where an auxiliary isoscalar gauge field is introduced, 
and a four component isovector is assumed to be transformed as a Lorentz 4-vector. 
However, the above analysis clarifies another problem; 
the extended Lorentz invariance is still broken by the mass term of the $\phi$ boson (\ref{MassTermForPhi}),  
even though the dispersion relation for each component is relativistic including unphysical modes.  
Whereas the components corresponding to the physical states have the unique mass $m_Y$, 
the propagator 
$\langle{\rm T}\phi_{\mu\nu}(x)\phi_{\rho\sigma}(x')\rangle$, which is calculable by the path integral method,  
is not covariant under the extended Lorentz transformation, 
since the unphysical states with masses $m_\phi=0, \frac{m_Y}{\sqrt{2}}$ also contribute to virtual processes.  

The effective Lorentz invariance would follow, if the mass term (\ref{MassTermForPhi}) was 
$-m_Y^2Y^{\mu\alpha}Y_{\mu\alpha}/2$, which is invariant under the extended Lorentz transformation.
However, this modification will be inappropriate for the picture of the Goldstone boson as the U(1) gauge field, 
since then the U(1) gauge field would become massive.
A consistent method to remove Lorentz violation from the effective theory 
is to separate the isospin rotation from the original Lorentz transformation by 
introducing a vierbein, and to transfer the isospin rotation to the local Lorentz transformation.
This improvement implies that one uses the extended gauge field as the local Lorentz frame vectors, 
as explained in the next section.

\section{Quantum vierbein\label{QV}}
The Lorentz invariance of the effective theory examined in the previous two sections reveals that  
the mass term of the symmetric tensor bosin $\phi_{\mu\nu}$ violates relativity in the extended sense. 
This section begins with introducing the vierbein into the same model to see how that point is modified. 

The extended gauge field $Y_{\mu\alpha}$ in the Minkowski spacetime is expressible in a local Lorentz frame as
\begin{equation}
Y_{\mu\alpha}=e_\mu{}^\beta Y_{\beta\alpha},
\label{LLY}
\end{equation}
with the help of the vierbein $e_\mu{}^\alpha$, 
where the first index of $Y_{\beta\alpha}$ represents a local Lorentz index, while the second is of the extended isospin, 
though both are expressed by the common greek letters.   
The arguments in the previous two sections will be applicable to $Y_{\beta\alpha}$  
by replacing a Lorentz transformation with a local Lorentz transformation.  
Owing to the vierbein, the expression (\ref{LLY}) becomes valid also 
in the curved spacetime with the metric $g_{\mu\nu}=e_\mu{}^\alpha e_{\nu\alpha}$.

Since the vierbein $e_\mu{}^\alpha$ has the same degree of freedom as $Y_{\mu\alpha}$,  
it serves to represent all the freedom of the gauge field, if every component of the vierbein is auxiliary. 
We hereafter forget the Einstein gravity for a while, 
and assume that all the components of the vierbein $e_{\mu\alpha}$ are auxiliary fields in a flat Minkowski spacetime. 
Considerations on the influence of the classical action of gravity is postponed until Sec.\ref{RGR}. 
Then, we will obtain the expression: 
\begin{equation}
Y_\mu{}^\alpha=e_\mu{}^\alpha\bar{\phi}, 
\label{Y=ePhi}
\end{equation}
where $\bar{\phi}$ is a scalar field of mass dimension one with the vacuum expectation value 
$\langle\bar{\phi}\rangle=\phi_0$,  
which is indispensable due to the normalization condition for the vierbein,  
$e^\mu{}_\alpha e_\mu{}_\beta=\eta_{\alpha\beta}$.  
Inversely, we can define a quantum version of the vierbein by 
\begin{equation}
\hat{e}_\mu{}^\alpha:=Y_\mu{}^\alpha/\bar{\phi}.
\label{e=Y/Phi}
\end{equation}
The Minkowski metric $\eta_{\mu\nu}$ is reproduced as the vacuum expectation value of the quantum metric 
\begin{equation}
\hat{g}_{\mu\nu}=\hat{e}_\mu{}^\alpha \hat{e}_{\nu\alpha}, \quad 
\langle\hat{g}_{\mu\nu}\rangle=\eta_{\mu\nu}.
\end{equation}
The contravariant quantum metric $\hat{g}^{\mu\nu}$ is given in terms of the inverse quantum vierbein  
\begin{equation}
\hat{g}^{\mu\nu}:=\eta^{\alpha\beta}\hat{e}^\mu{}_\alpha\hat{e}^\nu{}_\beta,  
\label{QuantumMetric}
\end{equation}
where the inverse quantum vierbein, satisfying 
$\hat{e}^\mu{}_\alpha\hat{e}_\mu{}^\beta=\eta_\alpha{}^\beta$ as well as 
$\hat{e}^\mu{}_\alpha\hat{e}_\nu{}^\alpha=\eta^\mu{}_\nu$,  
is obtainable by the definition  
\begin{equation}
\begin{array}{cc}
\hat{e}^\mu{}_\alpha:=-\displaystyle\frac{1}{3!\hat{e}}\epsilon^{\mu\nu\rho\sigma}\epsilon_{\alpha\beta\gamma\delta}
\hat{e}_\nu{}^\beta \hat{e}_\rho{}^\gamma \hat{e}_\sigma{}^\delta,&
\hat{e}=\det \hat{e}_\mu{}^\alpha,
\end{array}
\label{Def_InverseVierbein}
\end{equation}
where $\epsilon^{\mu\nu\rho\sigma}$ is the totally antisymmetric tensor with convention $\epsilon^{0123}=1$.  
The extended isospin space has been assumed to have the Minkowski metric $\eta_{\alpha\beta}$.   
In terms of the gauge field, Eq.(\ref{Def_InverseVierbein}) is expressible as 
\begin{equation}
\begin{array}{cc}
Y^\mu{}_\alpha=-\displaystyle\frac{\bar{\phi}^2}{3!\vert Y\vert}\epsilon^{\mu\nu\rho\sigma}\epsilon_{\alpha\beta\gamma\delta}
Y_\nu{}^\beta Y_\rho{}^\gamma Y_\sigma{}^\delta,
&\vert Y\vert:=\det Y_\mu{}^\alpha, 
\end{array}
\label{Contravariant_Y}
\end{equation}
from which we have the relations 
\begin{equation}
Y_{\mu\alpha}=\hat{g}_{\mu\nu}Y^\nu{}_\alpha, \quad 
\hat{g}_{\mu\nu}=Y_\mu{}^\alpha Y_{\nu\alpha}/\bar{\phi}^2, \quad
\bar{\phi}^2=Y^\mu{}_\alpha Y_\mu{}^\alpha/4.
\end{equation}
Since $\hat{g}_{\mu\nu}$ is defined by the gauge field, it changes under the SU(2)$\times$U(1) gauge transformation as  
\begin{equation}
\langle\delta_\theta\hat{g}_{\mu\nu}\rangle=(\partial_\mu\theta_\nu+\partial_\nu\theta_\mu)/\phi_0, \quad
\delta_\theta Y_\mu{}^\alpha=(\partial_\mu\theta^0, \partial_\mu\bm{\theta}+g\bm{\theta}\times \bm{Y}_\mu),
\end{equation}
which is the same as the infinitesimal coordinate transformation of the classical Minkowski metric $\eta_{\mu\nu}$
under the identification $\delta x^\mu=-\theta^\mu/\phi_0$. 
This observation suggests that if we formulate spontaneous Lorentz violation of the massive SU(2) gauge theory 
to be in general coordinate invariant in the quantum spacetime defined by $\hat{g}_{\mu\nu}$, 
the broken gauge symmetry will recover and the massive gauge bosons will return massless. 
In fact, the mass term of the gauge boson is transformed under the quantum spacetime into the mass term of the scalar boson, 
\begin{equation}
\frac{m_Y^2}{2}\hat{g}^{\mu\nu}\bm{Y}_\mu\cdot\bm{Y}_\nu=-\frac{3}{2}m_Y^2\bar{\phi}^2,
\end{equation} 
which is Lorentz invariant in the extended sense. 

The formulas in differential geometry are not altered whether the vierbein and the metric tensor are composed of the gauge field or not. 
In the quantum geometry, the covariant derivative for the general coordinate transformation 
and that for the local Lorentz transformation are obtainable by replacing $g_{\mu\nu}$ and $e_{\mu\alpha}$ 
by $\hat{g}_{\mu\nu}$ and $\hat{e}_{\mu\alpha}$, respectively: 
\begin{eqnarray}
\hat{\nabla}_\mu V_\nu&=&\partial_\mu V_\nu-\hat{\Gamma}^\rho_{\mu\nu} V_\rho,
\quad
\hat{\Gamma}^\rho_{\mu\nu}=\displaystyle\frac{1}{2}\hat{g}^{\rho\sigma}
(\partial_\mu\hat{g}_{\sigma\nu}+\partial_\nu\hat{g}_{\sigma\mu}-\partial_\sigma\hat{g}_{\mu\nu}),\\
\hat{\nabla}_\mu V^\alpha&=&\partial_\mu V^\alpha+\hat{\omega}_\mu{}^\alpha{}_\beta V^\beta,\quad 
\hat{\omega}_{\mu\alpha\beta}=\hat{e}^\nu{}_\alpha\hat{\nabla}_\mu\hat{e}_\nu{}_\beta   
=\hat{e}^\nu{}_\alpha(\partial_\mu\hat{e}_{\nu\beta}-\hat{\Gamma}^\rho_{\mu\nu}\hat{e}_{\rho\beta}).\label{LLC}
\end{eqnarray}
In accordance with the definition by Weinberg\cite{Weinberg1}, 
the covariant derivative for $\hat{e}_\mu{}^\alpha$ in (\ref{LLC})   
does not contain the local Lorentz connection $\hat{\omega}_{\mu\alpha\beta}$. 

The covariant derivative for a spinor field requires the spin connection $\hat{\omega}_\mu$. 
In the case of a Weyl spinor $\varphi$, it is expressible in terms of $\hat{\omega}_{\mu\alpha\beta}$ as
\begin{equation}
\hat{\nabla}_\mu\varphi=(\partial_\mu+\hat{\omega}_\mu)\varphi, \quad 
\hat{\omega}_\mu=\frac{1}{8}\hat{\omega}_{\mu\alpha\beta}\bar{\sigma}^{\alpha\beta}, \quad 
 \bar{\sigma}^{\alpha\beta}:=\sigma^\alpha\bar{\sigma}^\beta-\sigma^\beta\bar{\sigma}^\alpha
\end{equation}
where the four component Pauli matrices $\sigma^\alpha$ and $\bar{\sigma}^\alpha$ are defined by
\begin{equation}
\begin{array}{cccc}
\sigma^\alpha=(1, \bm{\sigma}),&\bar{\sigma}^\alpha=(1, -\bm{\sigma}).
\end{array}
\end{equation}
\section{Extra Gravity}
This section shows that the introduction of quantum geometry eliminates from the Lagrangian of the gauge field Eq.(\ref{Lag_Y}) 
the breakdown of extended Lorentz invariance.   
Actually, the mass term and the fourth order term in (\ref{Lag_Y}), 
which violate relativity in the extended sense, 
turn into the mass term and the self-interaction term for the scalar field $\bar{\phi}$: 
\begin{equation}
\frac{m_Y^2}{2}\bm{Y}^\mu\cdot\bm{Y}_\mu=
-\frac{3}{2}m_Y^2\bar{\phi}^2,\quad 
-\frac{g^2}{4}(\bm{Y}^\mu\times\bm{Y}^\nu)\cdot(\bm{Y}_\mu\times\bm{Y}_\nu)=-\frac{3}{2}g^2\bar{\phi}^4.  
\end{equation}
We will see that the quantum geometry also makes the third order term in Eq.(\ref{Lag_Y}) harmless to the emergent Lorentz invariance. 
The remaining second order term can be made relativistic in the extended sense as follows. 

As done in Sec.\ref{GP}, we introduce an auxiliary vector field $Y_{\mu0}$ 
to extend the isospin to be of four components, and a constant timelike 4-vector $\delta^\alpha$ for a local Lorentz violation. 
In a local Lorentz frame in which the spacial components of a timelike 4-vector vanish,  
$\delta^\alpha=(\delta,0,0,0)$, 
the second order term can be decomposed into the form 
\begin{eqnarray}
{\cal L}_Y^{(2)}&=&-\frac{1}{4}\bm{Y}^{\mu\nu}\cdot\bm{Y}_{\mu\nu}={\cal L}_{YG}+{\cal L}_A,\\
{\cal L}_{YG}&=&\frac{1}{4}Y^{\mu\nu\alpha}Y_{\mu\nu\alpha}, \quad 
Y_{\mu\nu\alpha}:=\partial_\mu Y_{\nu\alpha}-\partial_\nu Y_{\mu\alpha},\\ 
{\cal L}_A&=&-\frac{1}{4}F^{\mu\nu}F_{\mu\nu}, \quad 
F_{\mu\nu}:=\partial_\mu A_\nu-\partial_\nu A_\mu,\\
A_\mu&=&Y_{\mu0}-\langle Y_{\mu0}\rangle.
\end{eqnarray}
The contravariant tensors are assumed to be defined by the covariant ones with the help of the quantum metric, for example,   
$\bm{Y}^{\mu\nu}=\hat{g}^{\mu\rho}\hat{g}^{\nu\sigma}\bm{Y}_{\rho\sigma}$,  
although the quantum geometry makes no difference to the second order approximation of ${\cal L}_Y^{(2)}$. 
Following the argument in Sec.\ref{GP}, we hereafter treat $Y_{\mu\alpha}$ and $A_\mu$ as independent.  
The Lagrangian ${\cal L}_{YG}$ turns into that of quantum gravity by replacing $Y_{\mu\alpha}$ with $\hat{e}_{\mu\alpha}\bar{\phi}$. 
For a small quantum oscillation,  
$\hat{e}_{\mu\alpha}=\eta_{\mu\alpha}+\omega_{\mu\alpha}+{\cal O}(\omega^2)$, 
${\cal L}_{YG}$ becomes up to the quadratic order, 
\begin{equation}
{\cal L}_{YG}=\frac{\phi_0^2}{2}\left[\partial^\rho\omega^{\mu\alpha}\partial_\rho\omega_{\mu\alpha}
-\frac{1}{2}\partial^\rho\omega\partial_\rho\omega\right]
+\partial^\rho\bar{\phi}\partial_\rho\bar{\phi}+\partial^\rho\phi\partial_\rho\phi
-\frac{1}{2}\partial\cdot Y^\alpha\partial\cdot Y_\alpha, 
\label{L_YG}
\end{equation}
where total divergence terms are discarded. 
Since $\partial\cdot Y_\alpha=\partial_\alpha\phi+\phi_0(\partial^\rho\omega_{\rho\alpha}-\frac{1}{2}\partial_\alpha\omega)$ 
in the first order approximation, 
where $\phi=\sqrt{\hat{e}}\bar{\phi}$, and $\omega:=\omega^\rho{}_\rho$,  
the gauge fixing condition 
\begin{equation}
\partial^\rho\omega_{\rho\alpha}-\frac{1}{2}\partial_\alpha\omega=(\sqrt{2}-1)\partial_\alpha\phi/\phi_0 
\label{GaugeCondition}
\end{equation}
cancels the last two terms in the right-hand side of Eq.(\ref{L_YG}).
Then, we obtain in this gauge the following expression:
\begin{equation}
{\cal L}_Y=\frac{\phi_0^2}{8}\left[\partial^\rho \hat{h}^{\mu\nu}\partial_\rho \hat{h}_{\mu\nu}
-\frac{1}{2}\partial^\rho \hat{h}\partial_\rho \hat{h}\right]
+\frac{\phi_0^2}{2}\partial^\rho\epsilon^{\mu\nu}\partial_\rho\epsilon_{\mu\nu}
+{\cal L}_Y^{(3)}-\frac{1}{4}F^{\mu\nu}F_{\mu\nu}
+\partial^\rho\bar{\phi}\partial_\rho\bar{\phi}-\frac{3}{2}m_Y^2\bar{\phi}^2
-\frac{3}{2}g^2\bar{\phi}^4,
\label{QA}
\end{equation}
where ${\cal L}_Y^{(3)}=g\partial_\mu\bm{Y}_\nu\cdot(\bm{Y}^\mu\times\bm{Y}^\nu)$, 
$\hat{h}_{\mu\nu}=2\omega_{(\mu\nu)}=\omega_{\mu\nu}+\omega_{\nu\mu}$, 
$\epsilon_{\mu\nu}=\omega_{[\mu\nu]}=(\omega_{\mu\nu}-\omega_{\nu\mu})/2$, and $\hat{h}=\hat{h}^\rho{}_\rho$.   

For $\epsilon_{\mu\nu}=0$ with $\phi=\phi_0$, 
the gauge fixing condition (\ref{GaugeCondition}) reduces to $\partial^\rho \hat{h}_{\rho\mu}-\frac{1}{2}\partial_\mu \hat{h}=0$,  
which is the same as that of the harmonic coordinates for the gravitational wave $\delta\hat{g}_{\mu\nu}=\hat{h}_{\mu\nu}$\cite{Weinberg2}.    
The first part for $\hat{h}_{\mu\nu}$ in Eq.(\ref{QA}) is the same as the Lagrangian 
of gravitational waves in Einstein's gravity, 
provided that the gravitational constant $8\pi G$ is now replaced with $\phi_0^{-2}$.
The terms representing the Lagrangian of the scalar field $\bar{\phi}$ 
would turn into renormalization terms, 
if we had prepared beforehand the Lagrangian for $\bar{\phi}$. 
In this form of Lagrangian (\ref{QA}), only $\bar{\phi}$ develops the vacuum expectation value. 

Our remaining task is to evaluate ${\cal L}_Y^{(3)}$, which reduces in the same order of approximation to 
\begin{equation}
\hat{e}{\cal L}_Y^{(3)}=g\epsilon_{abc}\left[
-\hat{h}^\rho{}_a(\partial_\rho\omega_{bc}-\partial_b\omega_{\rho c})
+\omega^\rho{}_a(\partial_\rho\omega_{bc}-\partial_b\omega_{\rho c})
-\frac{1}{2}\omega\partial_a\omega_{bc}\right].
\label{L_3}
\end{equation}
The contribution from ${\cal L}_Y^{(3)}$ to the linear equation for $\omega_{\mu\nu}$ can be extracted 
by taking the variation of Eq.(\ref{L_3}):
\begin{equation}
\delta(\hat{e}{\cal L}_Y^{(3)})=g\epsilon_{abc}\left[
\begin{array}{l}
-\delta \hat{h}^{\mu\nu}\left(\eta_{\nu a}(\partial_\rho\omega_{bc}-\partial_b\omega_{\rho c})
-\frac{1}{4}\eta^{\mu\nu}\partial_a\omega_{bc}\right)\\
+\delta\omega^\mu{}_a
\left(\partial_\mu\omega_{bc}+\partial_b(\omega_{c\mu}-\omega_{\mu c})
+\eta_{\mu b}(\partial^\rho\omega_{\rho c}-\frac{1}{2}\partial_c\omega)\right)
\end{array}\right].
\label{DL_3}
\end{equation}
The first part proportional to $\delta\hat{h}^{\mu\nu}$ corresponds to the source term for the linear equation of $\omega_{\mu\nu}$, 
which contributes to the energy-momentum tensor of $\omega_{\mu\nu}$.  
The second part proportional to $\delta \omega^\mu{}_a$, which contributes to the free equation of motion, 
vanishes, if $\omega_{\mu\nu}$ is symmetric and satisfies the gauge condition 
$\partial^\rho\omega_{\rho \alpha}-\frac{1}{2}\partial_\alpha\omega=0$. 
Therefore, ${\cal L}_Y^{(3)}$ does not affect the free equation of motion for the extra graviton. 
We will see in Sec.\ref{RQP} that the antisymmetric tensor boson decouples from fermions by 
the local Lorentz transformation for spinor fields.   
Then, ${\cal L}_Y$ is effectively Lorentz invariant in the extended sense, even after spontaneous Lorentz violation. 
\section{Relativistic quasiparticles\label{RQP}}
We now turn our attention to the fermions coupled with the gauge field $\bm{Y}_\mu$, 
and examine the relativity of quasiparticles and their interactions.
In perturbation theory, the effective Lagrangian is separated into the free part and the interaction part.  
This section shows that the hypothesis  
$\langle Y_{\mu\alpha}\rangle=\phi_0\eta_{\mu\alpha}$ suffices for the Lorentz invariance of the free part,  
whereas the quantum geometry by the gauge field is indispensable for the Lorentz invariance of interactions. 

We assume here that the Lagrangian of fermions ${\cal L}_F$ is the sum of those for left-handed Weyl doublets,  
${\cal L}_F=\sum_{i=1}^N{\cal L}_W^{(i)}$. 
The fermion current $\bm{j}_\mu$ in the equation of motion (\ref{EqM_Y}) is then the sum of the currents  
from the doublets: $\bm{j}_\mu=\sum_{i=1}^N\bm{j}_\mu^{(i)}$, 
where $N$ should be even, if our model cancels the Witten's global SU(2) anomaly \cite{Witten,CDG,kiskis,Bar}. 

We consider one of the Weyl doublets, and give the Lagrangian ${\cal L}_W$ in the quantum spacetime 
described by the quantum vierbein $\hat{e}_\mu{}^\alpha$:   
\begin{equation}
\begin{array}{cccc} 
{\cal L}_W=({\cal L}_\varphi+{\cal L}_\varphi^\dagger)/2,&
{\cal L}_\varphi=\hat{e}^\mu{}_\alpha\varphi^\dagger\bar{\sigma}^\alpha iD_\mu\varphi, &
D_\mu=\partial_\mu+\frac{1}{8}\hat{\omega}_{\mu\beta\gamma}\bar{\sigma}^{\beta\gamma}
-i\frac{g}{2}\bar{\rho}_\alpha Y_\mu{}^\alpha, 
\end{array}
\label{L_W}
\end{equation}
where the four component isospin matrices, $\rho^\alpha=(1, \bm{\rho})$, $\bar{\rho}^\alpha=(1, -\bm{\rho})$, are introduced, 
and the gauge field is understood as $Y_\mu{}^\alpha=\hat{e}_\mu{}^\alpha\bar{\phi}$. 
The local Lorentz connection $\hat{\omega}_{\mu\beta\gamma}$ has already been given by Eq.(\ref{LLC}).  
We have assumed in Eq.(\ref{L_W}) an SU(2)$\times$U(1) gauge interaction including an auxiliary field $Y_{\mu0}$.  
The coupling constant $g'$ for $Y_{\mu0}$ has been set equal to $g$.  
Another choice of  $g'$ would affect the electric charges of quasifermions, as in the case of the standard electroweak theory. 

For small quantum oscillations, $\hat{e}_{\mu\alpha}=\eta_{\mu\alpha}+\omega_{\mu\alpha}$ and 
$\bar{\phi}=\phi_0+\sigma_X/\sqrt{2}$, ${\cal L}_W$ reduces to  
\begin{eqnarray}
{\cal L}_W&=&{\cal L}_W^0+{\cal L}_W^1+\cdots,\\
{\cal L}_W^0&=&\varphi^\dagger\left[\bar{\sigma}^\mu i\partial_\mu+\frac{m}{2}\bar{\rho}\cdot\bar{\sigma}\right]\varphi,
\quad m:=g\phi_0,\label{L_w0}\\
{\cal L}_W^1&=&m\omega_{\mu\nu}j^{\mu\nu}-\omega_{(\mu\nu)}\bar{K}^{\mu\nu}
+\frac{g}{\sqrt{2}}\sigma_Xj^\alpha{}_\alpha,\label{IntLag}\\
j^{\mu\nu}&=&\frac{1}{2}\varphi^\dagger\bar{\sigma}^\mu\bar{\rho}^\nu\varphi, \quad
K^{\mu\nu}=\varphi^\dagger\bar{\sigma}^\mu iD^\nu\varphi,  \quad
D_\mu=\partial_\mu-i\frac{m}{2}\bar{\rho}_\mu,
\label{FOL_w}
\end{eqnarray}
where $\bar{K}^{\mu\nu}$ is the real part of $K^{\mu\nu}$.
The derivation is given in the Appendix. 

We first examine the free equation of motion of a left-handed doublet obtained from Eq.(\ref{L_w0}) 
in momentum representation:  
\begin{equation}
\left(\bar{\sigma}\cdot p+\frac{m}{2}\bar{\rho}\cdot\bar{\sigma}\right)\varphi_p=0, 
\label{EM_QP}
\end{equation}
which gives the dispersion relation,   
\begin{equation}
\begin{array}{cc}
\vert\bar{\sigma}\cdot p+\frac{m}{2}\bar{\rho}\cdot\bar{\sigma}\vert=p\cdot p\left[(p-\delta)\cdot(p-\delta)-m^2\right]=0,&
\delta^\mu=-m\eta^{\mu0}.
\end{array}
\label{DRQL}
\end{equation}
In terms of the helicity eigenstates $L$ and $R$, the eigenfunctions are expressible as 
\begin{eqnarray}
p^0&=&p,\quad\varphi_{\nu\bm{p}}=LL,\label{QFS1}\\
p^0&=&-p,\quad\varphi_{\bar{\nu}\bm{p}}=RR,\label{QFS2}\\
p^0&=&\omega-m,\quad\varphi_{e\bm{p}}=\lambda_+RL+\lambda_-LR,\label{QFS3}\\
p^0&=&-\omega-m,\quad\varphi_{\bar{e}\bm{p}}=-\lambda_-RL+\lambda_+LR,\label{QFS4}
\end{eqnarray}
where $p=|\bm{p}|$ and $\omega=\sqrt{p^2+m^2}$, and 
the coefficients $\lambda_\pm$ are given by
\begin{equation}
\lambda_\pm=\sqrt{\frac{1}{2}\left(1\pm\frac{p}{\omega}\right)}.
\label{Def_Lambda_pm}
\end{equation}
The first $L$ or $R$ of each direct product in the wave functions corresponds to the isospin-helicity eigenstate, 
while the second is to the ordinary spin-helicity eigenstate. 

We may call the solution (\ref{QFS1}) with energy $E_{\nu p}=p$ a ``quasineutrino" and 
(\ref{QFS2}) with energy $E_{\bar{\nu} p}=p$ a ``quasi-anti-neutrino," 
while the solution (\ref{QFS3}) with energy $E_{ep}=\omega-m$ a ``quasielectron", and 
(\ref{QFS4}) with energy $E_{\bar{e}p}=\omega+m$ a ``quasipositron,"
in accordance with the hole theory applied to the negative kinetic energy. 

The dispersion relation for the quasielectron has the form 
$(p-\delta)\cdot(p-\delta)=m^2$.
A  fermion obeying this dispersion relation is an ordinary relativistic fermion 
with kinetic 4-momentum $k^\mu=p^\mu-\delta^\mu$.    
The Lorentz violating term $\delta^\mu$ can be removed, if the fermion has a U(1) gauge interaction.   

The momenta $p^\mu$ and $k^\mu$, which are transformable to each other under gauge transformations, 
are both physical quantities in quantum mechanics.
Then, the difference of the two, $\delta^\mu$, will be also physical. 
However, this does not imply that $\delta^\mu$ is always detectable, or has a physical effect.
It will depend on the situation, or what experiment is done, 
although we usually expect that $\delta^\mu$ does not have any physical effects\cite{CK1,CK2}.  

The difference between the quasielectron and the Dirac electron deserves special attention, 
since the former has the only one spin state like a neutrino.
The form of the wave function of the quasielectron represents a superposition of 
a right-handed electron and a left-handed electron, despite starting from a left-handed doublet. 
The appearance of a right-handed quasielectron is understandable as follows. 
The equation of motion (\ref{EM_QP}) is expressible for a quasielectron in the form
\begin{equation}
\begin{array}{cc}
\bar{\sigma}\cdot (p-\delta)\varphi_{ep}=m\chi_{ep},&\chi_{ep}:=T\varphi_{ep},
\end{array}
\label{EqOfM_qe}
\end{equation}
where 
\begin{equation}
T=\frac{1}{2}(1+\bm{\rho}\cdot\bm{\sigma}).
\label{EM_QL}
\end{equation}
The $4\times4$ matrix $T$ is unitary, and exchanges the spin matrices and the isospin matrices:
\begin{equation}
\begin{array}{ccc}
T^\dagger=T^{-1}=T,&T\bm{\rho}T^{-1}=\bm{\sigma},&T\bm{\sigma}T^{-1}=\bm{\rho}.  
\end{array}
\label{T_rho_sigma}
\end{equation}
Operating $T$ on Eq.(\ref{EqOfM_qe}), we find that $\chi_{ep}$ satisfies the equation of motion 
\begin{equation}
\bar{\rho}\cdot (p-\delta)\chi_{ep}=m\varphi_{ep}.
\label{EqM_chi}
\end{equation}
It can be rewritten as 
\begin{equation}
\sigma\cdot (p-\delta)\chi_{ep}=m\varphi_{ep},
\label{RHWD}
\end{equation}
due to the relation, $p\cdot(\sigma-\bar{\rho})\chi_e=0$, obtainable from the property of the wave function  
$\chi_{e\bm{p}}=\lambda_+LR+\lambda_-RL$. 
Equation (\ref{RHWD}) shows that $\chi_e$ is a right-handed doublet. 

Incidentally, the right-handed electron is also interpretable  
as a three body bound state of Weyl fermions, 
\begin{equation}
\chi_e=T\varphi_e\simeq\frac{2}{k_1m}\langle\varphi\varphi^\dagger\varphi\vert e\rangle,
\label{3BS}
\end{equation}
where $k_1$ is a quadratically divergent constant in a perturbative estimation.  
It is derived by expanding the propagator of a Weyl doublet with respect to $\bar{M}$. 
Then, we have
\begin{equation}
\langle\varphi(x)\varphi^\dagger(x)\rangle=\displaystyle\int\frac{d^4p}{(2\pi)^4}\frac{i}{\bar{\sigma}\cdot p+\bar{M}}
=\frac{k_1}{2}M+\cdots, 
\label{WDP}
\end{equation}
where
\begin{equation}
\begin{array}{cc}  
M=\frac{g}{2}\sigma^\mu\bar{\rho}^\alpha\langle Y_{\mu\alpha}\rangle=mT,&
k_1=\displaystyle\int\frac{d^4p}{(2\pi)^4}\frac{i}{p^2+i\epsilon},
\end{array}
\label{Def_M}
\end{equation}
from which Eq.(\ref{3BS}) follows. 
In grand unified theories like an SU(5) model\cite{GG}, 
it is not unusual that a left-handed state and a right-handed state of the same particle belong to different multiplets. 

The interaction part of the Lagrangian (\ref{IntLag}), on the other hand, is expressible in the form:
\begin{eqnarray}
{\cal L}_W^1&=&-\omega_{(\mu\nu)}\bar{T}^{\mu\nu}
+\epsilon_{\mu\nu}\frac{m}{2}\varphi^\dagger\bar{\sigma}^\mu\bar{\rho}^\nu\varphi
-\omega_{(\mu\nu)}\varphi^\dagger\bar{\sigma}^\mu\varphi\delta^\nu
+\frac{g}{\sqrt{2}}\sigma_X\varphi^\dagger(1-T)\varphi,
\label{L_w^1}\\
T^{\mu\nu}&=&\varphi^\dagger\bar{\sigma}^\mu (i\partial^\nu-\delta^\nu)\varphi.
\end{eqnarray}
For the quasineutrino: $\varphi=\varphi_\nu$, $\delta^\mu=0$,  
${\cal L}_W^1$ reduces to
\begin{equation}
{\cal L}_W^1=-\frac{\hat{h}_{\mu\nu}}{2}\bar{T}^{\mu\nu}.
\label{L_nu^1}
\end{equation}
There remains only a gravitational interaction, since Eqs. (\ref{QFS1}) and (\ref{QFS2}) show that  
the wave functions of the quasineutrino are symmetric under the exchange of the spin and the isospin, 
from which $T\varphi_\nu=\varphi_\nu$ follows. 

In the case of the quasielectron: $\varphi=\varphi_e$, $\delta^\mu=-m\eta^{\mu0}$, 
the interaction Lagrangian (\ref{L_w^1}) is also expressible in a Lorentz invariant form:
\begin{eqnarray}
{\cal L}_W^1+\delta_L\left[-m\varphi_e^\dagger T\varphi_e\right]&=&-\frac{\hat{h}_{\mu\nu}}{2}\bar{T}_e^{\mu\nu}
+gA_\mu\varphi_e^\dagger\bar{\sigma}^\mu\varphi_e
-\frac{g}{\sqrt{2}}\sigma_X\varphi_e^\dagger\chi_e,\label{L_e^1}\\
gA_\mu&=&gY_{\mu0}-\langle gY_{\mu0}\rangle\simeq m\omega_{\mu0}+\eta_{\mu0}\frac{g}{\sqrt{2}}\sigma_X. 
\end{eqnarray}
The second term in the left-hand side of Eq.(\ref{L_e^1}) is a local Lorentz transformation of the quasielectron mass term:
\begin{equation}
\delta_L\left[-m\varphi_e^\dagger T\varphi_e\right]
=m\epsilon_{\mu\nu}\varphi_e^\dagger(\bar{\sigma}^\mu\eta^{\nu0}-\frac{1}{2}\bar{\sigma}^\mu\rho^\nu)\varphi_e, \quad 
\delta_L\varphi_e=\frac{1}{8}\epsilon_{\mu\nu}\bar{\sigma}^{\mu\nu}\varphi_e.
\label{LLTM}
\end{equation}
Although the interaction Lagrangian (\ref{L_e^1}) is expressed in terms of a left-handed doublet only, 
we can also find another local Lorentz transformation which reproduces the left-right symmetric amplitudes\cite{KN2}. 
The quasielectron has a gravitational interaction, a U(1) gauge interaction, 
and a scalar interaction coupling to the mass term of a quasielectron.
\section{Reproduction of General Relativity\label{RGR}}
After confirming the Lorentz invariance of the effective theory,  
we finally consider the phenomenological implications of the results obtained. 
We first assume that Lorentz symmetry is spontaneously broken in the present phase of the vacuum. 
Since the original massive SU(2) gauge bosons disappear in this phase, 
these bosons will not be identical with the weak bosons in the standard electroweak theory. 
Further, if the Einstein type action of gravity is already present even before spontaneous Lorentz violation, 
we should add the Lagrangian of gravity:
\begin{equation}
\sqrt{-g}{\cal L}^0_G=\frac{-1}{16\pi G_0}\sqrt{-g}R\simeq
\frac{1}{64\pi G_0}\left[\partial^\rho h^{\mu\nu}\partial_\rho h_{\mu\nu}
-\frac{1}{2}\partial^\rho h\partial_\rho h\right],
\label{Lag_EG^0}
\end{equation}
to  Eq.(\ref{QA}), and the interaction Lagrangian, 
\begin{equation}
{\cal L}^1_G=-\frac{h_{\mu\nu}}{2}\bar{T}^{\mu\nu},
\label{Lag_EG^1}
\end{equation}
to Eq.(\ref{IntLag}) under our approximation, where $G_0$ is the bare gravitational constant, 
and $\delta g_{\mu\nu} =h_{\mu\nu}$ is a small oscillation for the classical metric before spontaneous Lorentz violation. 

Since $\hat{h}_{\mu\nu}$ and $h_{\mu\nu}$ generate independent gravitational forces, 
the present Newton constant should be the sum of the two gravitational constants $G_0$ and $G_{\rm ex}=(8\pi\phi_0^2)^{-1}$:
\begin{equation}
G_N= G_0+\frac{1}{8\pi\phi_0^2}. 
\end{equation}
The ratio of the bare gravitational constant to the Newton constant is expressed by 
\begin{equation}
G_0/G_N=1-\frac{g^2M_P^2}{8\pi m^2},
\end{equation}
where $M_P$ is the Planck mass.

We first temporarily assume that $G_0=0$, which implies that the Yang-Mills gravity can approximate Einstein's gravity,  
although the coincidence of two theories beyond perturbation remains yet to be seen. 

Further, if the quasifermions are identified with leptons in this case, 
$m=g\phi_0$ should have the order of a charged lepton mass $m_\ell=m_e$, $m_\mu$, or $m_\tau$, 
and the coupling constant should be extremely small: $g\simeq m_\ell/M_P\ll1$.  
Therefore, the U(1) gauge field $A_\mu$ in Eq.(\ref{L_e^1}) will represent an extra electromagnetism, 
which interacts with leptons extremely weaker than the ordinary electromagnetism. 

As a possible option, we may regard $A_\mu$ as the real photon 
by assuming the coupling constant to be the electric charge of the electron, $g=e$.
Then, the charged lepton would in turn become extremely heavy, $m\simeq M_P$. 

In the case of $G_0\neq0$, on the other hand, 
our model can represent a unified theory of leptons interacting with a graviton, a photon, and a Higgs-like scalar boson, 
without any of the inconveniences found above.    
However, this case predicts that the bare gravity present before spontaneous Lorentz violation will be an extremely strong antigravity, 
$-G_0/G_N\simeq{\cal O}(M_P^2/m_\ell^2)$. 
This conclusion will not be avoidable, as long as the quasifermions are identified with real leptons, and the Goldstone photon with the real photon. 
We here do not enter into considerations on the quantum mechanical problems with antigravity, 
nor on the influence upon cosmology, including the big bang scenario of the Universe. 
\section{Summary}
We have shown that the effective theory emergent from spontaneous Lorentz violation 
of massive SU(2) gauge theory of Weyl doublets in four dimensions will be Lorentz invariant under the two hypotheses:  

1. The extended gauge field $Y_{\mu\alpha}$, including an auxiliary field $Y_{\mu0}$,  
develops vacuum expectation value $\langle Y_{\mu\alpha}\rangle=\eta_{\mu\alpha}\phi_0$.

2. The local Lorentz frame is provided by $Y_{\mu\alpha}$ according to the relation  
$Y_{\mu\alpha}=\hat{e}_{\mu\alpha}\bar{\phi}$.

Then, the massive SU(2) gauge bosons are turned into a massless tensor boson $\omega_{\mu\nu}$, a Goldstone photon $A_\mu$,  
and a massive scalar boson $\bar{\phi}$.  
As a natural consequence from the second hypothesis, an extra graviton 
appears as the symmetric part of the massless tensor boson. 
The correspondence between the Einstein gravity and the Yang-Mills gravity beyond perturbation remains yet to be examined. 

In particular, if we regard our model as a unified theory of gravity and electromagnetism interacting with leptons, 
the bare gravitational constant before Lorentz violation will be negative and very large, 
which predicts the existence of an extremely strong primordial antigravity.  
Discussions on possible problems which may arise from antigravity are also beyond the scope of this paper. 

\appendix
\section{Derivation of the first order interaction Lagrangian (\ref{IntLag})\label{A3}}
For small variations $\delta\hat{e}_\mu{}^\alpha$ and $\delta Y_\mu{}^\alpha$, 
the Lagrangian ${\cal L}_\varphi$ in (\ref{L_W}) changes as
\begin{equation}
\delta{\cal L}_\varphi=\delta\hat{e}^\mu{}_\alpha K^\alpha{}_\mu
+\frac{1}{8}\hat{e}^\mu{}_\alpha\delta\hat{\omega}_{\mu\beta\gamma}
\varphi^\dagger i\bar{\sigma}^\alpha\bar{\sigma}^{\beta\gamma}\varphi
+g\hat{e}^\mu{}_\alpha\delta Y_{\mu\beta}j^{\alpha\beta}.
\end{equation}
For the real part of ${\cal L}_\varphi$, we have 
\begin{equation}
\delta{\cal L}_W=\delta\hat{e}^\mu{}_\alpha \bar{K}^\alpha{}_\mu
+\frac{1}{4}\epsilon^{\alpha\beta\gamma\delta}\hat{e}^\mu{}_\alpha\delta\hat{\omega}_{\mu\beta\gamma}
\varphi^\dagger\bar{\sigma}_\delta\varphi
+g\hat{e}^\mu{}_\alpha(\delta\hat{e}_{\mu\beta}\bar{\phi}+\eta_{\mu\beta}\delta\bar{\phi})j^{\alpha\beta},
\label{RealPartOfDLw}
\end{equation}
where the following identity has been used: 
\begin{equation}
\bar{\sigma}^\alpha\sigma^\beta\bar{\sigma}^\gamma=\eta^{\alpha\beta}\bar{\sigma}^\gamma
-\eta^{\alpha\gamma}\bar{\sigma}^\beta+\eta^{\beta\gamma}\bar{\sigma}^\alpha-i\epsilon^{\alpha\beta\gamma\delta}\bar{\sigma}_\delta. 
\label{sigma_identity}
\end{equation}
From the definition of the local Lorentz connection (\ref{LLC}), we obtain for $\hat{e}_{\mu\alpha}=\eta_{\mu\alpha}+\omega_{\mu\alpha}$: 
\begin{equation}
\epsilon^{\alpha\beta\gamma\delta}\hat{e}^\mu{}_\alpha\delta\hat{\omega}_{\mu\beta\gamma}
=\epsilon^{\alpha\beta\gamma\delta}\partial_\alpha\omega_{\beta\gamma}+\cdots.
\end{equation}
Then, ignoring a total divergence term, we have
\begin{equation}
\frac{1}{4}\epsilon^{\alpha\beta\gamma\delta}\hat{e}^\mu{}_\alpha\delta\hat{\omega}_{\mu\beta\gamma}
\varphi^\dagger\bar{\sigma}_\delta\varphi
=-\frac{1}{4}\epsilon^{\alpha\beta\gamma\delta}\omega_{\alpha\beta}\partial_\gamma(\varphi^\dagger\bar{\sigma}_\delta\varphi)+\cdots. 
\label{dOmega_J}
\end{equation}
On the other hand, $K^{\mu\nu}$ can be rewritten in the form 
\begin{equation}
K^{\mu\nu}=\varphi^\dagger\bar{\sigma}^\mu iD^\nu\varphi
=\frac{1}{2}\varphi^\dagger\bar{\sigma}^\mu(\sigma^\nu\bar{\sigma}^\rho+\sigma^\rho\bar{\sigma}^\nu) iD_\rho\varphi
=\frac{1}{2}\varphi^\dagger\bar{\sigma}^\mu\sigma^\rho\bar{\sigma}^\nu iD_\rho\varphi,
\end{equation}
by using the equation of motion, $\bar{\sigma}^\rho iD_\rho\varphi=0$. 
Then, from the identity (\ref{sigma_identity}), we have
\begin{equation}
K^{[\mu\nu]}=-\frac{1}{2}\epsilon^{\mu\nu\rho\sigma}\varphi^\dagger\bar{\sigma}_\sigma iD_\rho\varphi. 
\end{equation}
Therefore, the real part of $K^{[\mu\nu]}$ reads
\begin{equation}
\bar{K}^{[\mu\nu]}=-\frac{1}{4}\epsilon^{\mu\nu\rho\sigma}\partial_\rho(\varphi^\dagger\bar{\sigma}_\sigma\varphi). 
\end{equation}
Accordingly, Eq.(\ref{dOmega_J}) is expressible as
\begin{equation}
\frac{1}{4}\epsilon^{\alpha\beta\gamma\delta}\hat{e}^\mu{}_\alpha\delta\hat{\omega}_{\mu\beta\gamma}
\varphi^\dagger\bar{\sigma}_\delta\varphi
=\omega_{\mu\nu}\bar{K}^{[\mu\nu]}+\cdots.
\end{equation}
Since $\delta\hat{e}^\mu{}_\alpha=-\omega_\alpha{}^\mu$, Eq. (\ref{RealPartOfDLw}) reduces to 
\begin{equation}
\delta\bar{{\cal L}}_W=-\omega_{\mu\nu}\bar{K}^{(\mu\nu)}
+g\bar{\phi}\omega_{\mu\nu}j^{\mu\nu}+g\delta\bar{\phi}j^\alpha{}_\alpha+\cdots, 
\label{ReducedRealPartOfDLw}
\end{equation}
which is identical with Eq.(\ref{IntLag}) in the first order approximation, $\delta\bar{\phi}=\sigma_X/\sqrt{2}$.

\end{document}